\documentclass[manuscript]{aastex}

\usepackage{emulateapj5}

\slugcomment{Accepted by the Astrophysical Journal Letters}

\shorttitle{Impact of Reionization on Dwarf Galaxies}
\shortauthors{Grebel \& Gallagher}

\begin{document}

\title{Impact of Reionization on the Stellar Populations of  
    Nearby Dwarf Galaxies}

\author{Eva K. Grebel} 
\affil{Astronomical Institute, University of Basel, Venusstrasse 7,
CH-4102 Binningen, Switzerland}
\email{grebel@astro.unibas.ch}
    
\and

\author{John S. Gallagher, III}
\affil{University of Wisconsin, Department of Astronomy, 55343 Sterling,
475 N.\ Charter St., Madison, WI 53706-1582}
\email{jsg@astro.wisc.edu}

\begin{abstract}
Cold dark matter models for galaxy formation predict that low-mass
systems will be the first sites of star formation. As these objects
have shallow gravitational potential wells, the subsequent growth of
their stellar populations may be halted by heating and gas loss due to
reionization. This effect has been suggested to have
profoundly influenced properties of
present-day dwarf galaxies, including their stellar populations and
even survival as visible galaxies. In this Letter we draw on results
from quantitative studies of Local Group dwarf galaxy star formation
histories, especially for Milky Way satellites, to show that no clear
signature exists for a widespread evolutionary impact from
reionization.  All nearby dwarf galaxies studied in sufficient detail
contain ancient populations indistinguishable in age from the oldest 
Galactic globular clusters.  Ancient star formation activity proceeded
over several Gyr, and some dwarf spheroidal galaxies even experienced
fairly continuous star formation until just a few Gyr ago.
Despite their uniformly low masses, their star formation histories
differ considerably.  The evolutionary histories of nearby dwarf galaxies
appear to reflect influences from a variety of local processes
rather than a dominant effect from reionization.
\end{abstract}


\section{Introduction}

When did galaxies, or protogalactic fragments, first form 
stars?  When did they undergo their first major episodes of
star formation that produced the old Population II stars that we can still
observe today?  Did all of the early substructures begin to form stars
at the same time, or were there considerable differences from object to
object?  Did re-ionization squelch subsequent star formation in low-mass
substructures (Efstathiou 1992), accounting for the lower than predicted
number of Local Group satellite galaxies (e.g., Bullock, Kravtsov, \& Weinberg
2000; Somerville 2002; Benson et al.\ 2002, 2003; Tully et al. 
2002)?  How do the early star
formation histories that we can infer from their stellar fossil record fit
in with standard cold dark matter (CDM) paradigm?

Theory predicts that the earliest stars formed in low 
mass systems at $z \sim
30$ (e.g., Barkana \& Loeb 2001), a redshift that is not accessible 
for study with the current observational tools 
(e.g., Fan et al.\ 2001, 2003; Hu et al.\ 2002; Kodaira et al.\
2003; Pell\'o et al.\ 2004).  An alternative
approach concentrates on the local fossil record contained in those
present-day galaxies that we can study in the greatest detail.  
The highest accuracy in age-dating stellar populations can be reached
where stellar populations are resolved into individual stars down
to and below the strongly age-sensitive location of the main-sequence
turn-off (MSTO). 
This limits us to nearby galaxies.
The Milky Way and its satellites are sufficiently close and uncrowded
to make stars below the oldest MSTOs accessible 
with relative ease, and to break the age-metallicity degeneracy for
old, low-mass stars spectroscopically (see Grebel 1997; Grebel,
Gallagher, \& Harbeck 2003, hereafter GGH03; Cole et al.\ 2004). 
The dwarf satellites of the Milky Way offer 
the additional advantage of being close to the mass scales predicted for 
primitive sites hosting the first star formation in CDM models.

While Population III stars were presumably very massive, shortlived objects,
stars at the low-mass end of the subsequent, ``second'' generation of early
star formation
may still exist today and can be observed as extremely metal-poor stars
(Mackey, Bromm, \& Hernquist 2003).  But these ``Population
II.5'' stars (Mackey et al.) 
are rare and to date have only been detected in
the Milky Way (e.g., Chiba \& Beers 2000; Christlieb et al.\ 2002).
An additional difficulty lies in age-dating these 
individual stars. 

We therefore concentrate on stars belonging to Mackey et al.'s
(2003) ``third'' epoch of star formation, i.e., Population II stars.
In the Milky Way,
these populations comprise both field stars and old globular clusters.
We require that these ancient stars formed in sufficiently high numbers
that they (1) are still easily detectable today, and (2) produced
well-defined, measurable MSTOs usable for age dating, defining our
``oldest measurable episodes of star formation.'' We further focus 
on ancient stellar populations in nearby dwarf galaxies, which
have sufficiently low masses that their star formation 
should have been interrupted or possibly permanently squelched by the 
reionization of the universe (e.g., Thoul \& Weinberg 1996; 
Barkana \& Loeb 1999; Tassis et al. 2003). The
old stellar populations in dwarf galaxies thus should contain a 
record of a key phase in the early evolution of the universe (e.g., 
Gallagher \& Wyse 1994).

In this Letter we consider the star formation histories of nearby 
dwarf galaxies as derived from their resolved 
stellar populations. We compare levels and relative time scales for 
ancient star formation with the predicted effects 
of reionization based on time constraints from high-redshift quasars 
and Wilkinson Microwave
Anisotropy Probe (WMAP) measurements (Kogut et al.\ 2003; Spergel
et al.\ 2003).

\section{Reionization}

Reionization of the universe, which occurs via 
photoionization, can have a major impact on the evolution of low 
mass galaxies. Photo-heating of 
gas associated with a galaxy raises its temperature to the point where 
retention becomes an issue for small systems (Babul \& Rees 1992, 
Efstathiou 1992). The details 
of this process are very complex, with shielding and radiative 
transfer playing important roles.  Many models
predict that small galaxies should experience the bulk of their 
star formation {\it before} reionization (e.g., Ricotti, Gnedin, 
\& Shull 2002; Somerville 2002; Dekel \& Woo 
2003; Tassis et al.\ 2003; Susa \& Umemura 2004), and allow a restart 
of star formation only well after reionization. For example, Susa  
and Umemura (2004) confirm the suppression of star formation in galaxies 
with total masses of $M \leq$10$^9$~M$_{\odot}$ and collapse 
occurring at $z \leq$5 (see also Ferrara \& Tolstoy 2000).  Thus all low 
mass galaxies should have formed the bulk of their stars in about 
the first 1~Gyr after the Big Bang 
(see also Barkana \& Loeb 1999; Tassis et al.\ 2003).  We  
adopt a flat universe model 
with $\Omega_m =$0.27 and H$_0 =$ 71~km~s$^{-1}$~Mpc$^{-1}$
(Spergel et al.\ 2003).
   
While some models show star formation extending past reionization, 
the levels are low. We therefore can test the theory 
through measurements of stellar age distributions in Local Group 
galaxies. Susa \& Umemura (2004) did so and found qualitative 
agreement based on the star formation histories presented by 
Mateo (1998). However, new observations since have become available 
and so we can revisit this comparison with 
more quantitative data. Similarly, we wish to revisit the conclusions
presented by Gnedin (2000) on the basis of the Grebel (1999) and 
Mateo (1998) reviews for a nearly simultaneous drop in the star formation 
rates of Local Group dwarf spheroidal (dSph) galaxies
about 10 Gyr ago, which he associates with reionization.  
With typical masses of $10^7$~M$_{\odot}$ (Mateo 1998),
dSphs are the galaxies most likely to have been stripped of star-forming
material due to reionization.

To make this comparison, we need to constrain the redshift of 
reionization and find an associated time to compare with the timescales 
derived from stellar populations in dwarf galaxies. Further complications 
occur because reionization rates vary with location depending on 
the densities of matter and Lyman continuum photons. 
We assume that reionization typically occurs at $20 < z_{reion} <6.4$ 
with the upper bound coming from WMAP and the lower from 
quasar absorption line studies (Spergel et al.\ 2003; Fan et al.\
2003). With our choice of cosmology 
this translates to a time interval of about 0.2--0.9~Gyr after the 
Big Bang. 

\section{Relative ages of old populations in nearby galaxies}

Age-dating techniques for resolved old stellar populations
are summarized in Krauss \& Chaboyer (2003), including 
MSTO ages of stellar ensembles.  This is the most widely used
method since it is the observationally easiest technique.  The derivation
of {\em absolute}\ ages is model-dependent and
may result in age uncertainties of up to a few Gyr.  {\em Relative}\
age-dating techniques are easier to employ and allow one to reach a higher
differential accuracy (provided that the underlying assumptions hold, e.g.,
no variations in [$\alpha$/Fe] among the populations to be compared).
The relative dating techniques for old populations
typically rely on the position of the region around the MSTO
relative to other reference points in color space.
Alternatively, fiducials
(e.g., the mean ridge line of a coeval stellar
population in color-magnitude
space) registered to the MSTO region, or luminosity functions including
the MSTO are used (see Stetson, VandenBerg, \& Bolte 1996; 
Sarajedini, Chaboyer, \& Demarque 1997).

Old populations with ages $\ga 10$ Gyr have been detected in all Local
Group galaxies studied in sufficient detail and depth.  No galaxies 
{\em without} an old population have been found so far
(see Grebel 2000, 2001), but we are not yet certain
that all dwarf galaxies formed stars before the reionization epoch at 
z $\sim 6$.  

\subsection{Old dwarf spheroidal galaxies}

``Old'' dSphs comprise all dSphs that contain horizontal
branch stars but no intermediate-age
populations as traced by a red clump or a young blue main sequence.  The
four nearby Milky Way companions Draco, Ursa Minor, Sextans, Sculptor,
the M31 satellites Andromeda I--III, V, VI,
and the isolated dSphs Cetus and Tucana are members of this class
(e.g., Grebel 2000; Sarajedini et al.\ 2002; GGH03;
Harbeck, Gallagher, \& Grebel 2004).
When using horizontal branch (HB) {\em morphology}\ as a
relative age indicator, old dSphs seem to be 1 -- 2 Gyr younger than
the bulk of the old Galactic globular clusters (Harbeck et al.\ 2001), 
but age is not the only explanation for the second-parameter effect
(e.g., Salaris \& Weiss 2002) and so a younger age is not 
assured from the HB morphologies. 

MSTO techniques show the oldest populations in the nearby old dSph galaxies
Draco (Grillmair et al.\ 1998), Ursa Minor (Feltzing et al.\ 1999;
Mighell \& Burke 1999; 
Wyse et al.\ 2002), Sculptor (Monkiewicz et al.\ 1999), and Sextans
(Lee et al.\ 2003) to be indistinguishable in age from the 
oldest globular clusters in the Galactic  halo and bulge within 
measurement uncertainties of $\sim 1$ -- 1.5 Gyr.

Although these dSphs are dominated by old, metal-poor populations, they are not 
single-age, single-metallicity populations as found in globular clusters.
Spatial gradients in HB morphology indicate variations in the star formation
history as a function of position even in old dSphs (Harbeck et al.\ 2001).
All old dSphs exhibit significant metallicity spreads that may exceed 1 dex
in [Fe/H] (see GGH03).  
Hence the early star formation episodes must have been sufficiently 
long-lasting and a sufficient amount of the newly generated metals
must have been retained to lead to the observed enrichment.  The measured
elemental abundance ratios indicate both Type Ia and Type II supernova
enrichment (Shetrone et al.\ 2001; Tolstoy et al. 2003), 
requiring time scales of 1 -- 2 Gyr.
Chemical evolution modelling suggests even longer star formation time
scales of more than 4 Gyr (Ikuta \& Arimoto 2002).

\subsection{Mixed-age dwarf spheroidal galaxies}

Other dSph galaxies contain substantial intermediate-age populations.
Here the bulk of the stars was formed much less than 10 Gyr ago.  Still,
these mixed-age galaxies have old populations (including globular 
clusters) as old as the
oldest Galactic globular clusters (Sagittarius: Montegriffo et al.\ 1998; 
Layden \& Sarajedini 2000; Fornax: Buonanno et al.\ 1998;
Carina: Monelli et al.\ 2003; and  Leo\,II: Mighell \& Rich 1996).
Remarkably, in Fornax star formation ceased only as recently as 
100 -- 200 Myr ago (Grebel \& Stetson 1999; GGH03).
The available data indicate that star formation proceeded fairly
continuously; only Carina exhibits a clearly episodic star formation
history with a pause of several Gyr after the old population formed
(Smecker-Hane et al.\ 1994; Monelli et al.\ 2003).

\subsection{Other dwarf galaxies and satellites}

The Large Magellanic Cloud (LMC), the closest companion of the
Milky Way, has a mass of $5.3 \pm 1.0 \cdot 10^9$ M$_{\odot}$ (Alves \&
Nelson 2000)
and should therefore have been able to retain its gas after reionization.
Its old globular clusters are as old as ancient Galactic globulars 
(Olsen et al.\ 1998; Johnson et al.\ 1999).  
Deep studies of the LMC field population (e.g., Smecker-Hane et al.\ 
2002) also reveal ancient MSTOs 
and an overall smooth and continuous field star formation history.
The 
Small Magellanic Cloud (SMC) is a dwarf irregular (dIrr) galaxy with
$2\cdot 10^9$ M$_{\odot}$ (Westerlund 1997).
NGC 121, the only globular cluster in the SMC,
is $\sim 2$ -- 3 Gyr younger than the oldest Galactic
globular clusters ((e.g., Mighell et al.\ 1998; Shara et al.\ 1998).  
While RR Lyrae stars have been detected among SMC field stars (e.g.,
Graham 1975), field MSTO data with sufficient area coverage are still
lacking to address the question of pre-reionization star formation in
the SMC.

Less massive dIrr
and transition-type dwarf galaxies are located
at larger distances from the Milky Way.  Here the evidence for the
presence of old populations is usually based on the detection of HB stars
but the existing data do not reach the MSTO region, preventing
more accurate age dating.  Even so, the available data 
indicate star formation activity began early and 
proceeded smoothly over time with no major interruptions.

\section{Discussion}

With our choice of a flat universe cosmology and time scale 
for reionization (see \S2)
reionization was
complete $\sim 12.8$ Gyr ago, about 0.8~Gyr after the 
Big Bang.  If we adopt these
numbers (Fig.\ 1), then the low-mass
dSphs would have had only $\sim$1~Gyr after the
Big Bang to form stars before star formation would have been
inhibited by heating caused by reionization and feedback (e.g., Tassis
et al.\ 2003) or all gas would have effectively
been removed by photoevaporation (Barkana \& Loeb 1999). 

Converting the relative ages of old Galactic
globular clusters, against which the Galactic satellites are being compared,
into absolute ages is fraught with uncertainties.
However, Krauss \& Chaboyer (2003) find
a best-fit age of 13.4$\pm$2~ Gyr for the old Galactic globulars, 
which agrees well with the cosmological time frame established above.   
We conclude as follows: 
\begin{itemize}
\item
There is evidence for ancient star formation in the Milky
Way and its low-mass companions at times consistent with and required
by CDM models of dwarf galaxy formation, 
\item Within  the accuracy of relative MSTO age dating
($\sim 1$ Gyr), all dSphs for which such data are available
share a common epoch of ancient star formation with the
Milky Way (Grebel 2000, 2001), and 
\item As predicted by standard classes of 
CDM models for low-mass galaxies, no dwarf galaxies 
{\em without}\ ancient stellar populations are observed although 
the data are incomplete (e.g., in the SMC).
\end{itemize}

No clear indications exist for an imprint of the 
epoch of reionization on the star 
formation histories of nearby dSphs. Modern 
data show that star formation extended over $\sim 2$~Gyr 
or more even in the 
``old'' dSphs. The intermediate age systems for the most part had 
continuous star formation up until at least $\sim$5~Gyr in the past, 
corresponding to a redshift of $z \approx 0.5$; the
majority of stars in these systems formed at lower redshifts (see Fig.\ 1).
We also find galaxies with properties in 
common with the dSphs, the transition dIrr/dSph galaxies, where star 
formation continues into the present epoch (GGH03). 

The steep drop in star-formation activity in response to 
photoionization (e.g., Gnedin 2000
and Susa \& Umemura 2003), is not observed in these objects.
Only the Carina dSph displays 
the kind of gap in star formation that is expected if star formation were 
truncated during reionization and then restarted at low 
redshift, although this might be difficult to detect observationally
if the restart occurs much before z$\sim$1.

Apparently processes 
other than reionization shaped the star formation histories 
of most nearby dwarf 
galaxies. If reionization were a {\it dominant} influence on the evolution 
of stellar populations of dwarfs, then we might expect to see relatively 
uniform behavior among these systems.  This is clearly not the case. 
Gas-free dwarfs with older stellar populations preferentially are found 
near giant galaxies, while the dIrr/dSph galaxies with small gas reservoirs 
and the H\,{\sc I}-rich dIrrs avoid giants. Evidently local conditions are 
a major factor in the well known dwarf galaxy morphology--star formation 
history--environment connection 
(GGH03).  

Even among the gas-free dSphs, galaxies in similar locations can have 
radically different star formation histories (e.g., Grebel 1997
and \S3).  This complexity may be due to a variety of 
factors affecting the evolution of small galaxies; e.g., 
feedback, gas stripping, time dependence of gas infall, etc.\ (e.g., Somerville 
2002).  While reionization could 
be one such factor, it does not appear to have been sufficiently strong 
to produce coherent patterns of star formation among the 
Galaxy's dwarf companions. 

These results suggest that the gas supplies of nearby dwarfs were not 
so drastically reduced by reionization 
that their star formation
experienced  long-term disruptions.  
For example, gas already within dwarf galaxies may not be 
expelled by reionization, as suggested by analytic models 
(Somerville 2002, Benson et al. 2003).
Since the satellites of giant systems are expected to form 
near their host (Fukushige \& Makino 2001; Hayashi et al.\ 2003), they 
will be in a unique environment from early times, and thus are subject
to a variety of external influences, including tides 
in addition to the processes discussed 
earlier (Kravtsov, Gnedin, \& Klypin 2004).  
Another possibility arises from 
the uncertainties in the total masses of dSphs, i.e., whether they are 
or originally were 
much more massive than they appear from analyses based on their optical 
structures (e.g., Odenkirchen et al.\ 2001; Kleyna et al.\ 2001; Hayashi et 
al.\ 2003). If the Galactic dSphs are remnants of more 
massive galaxies (but see Klessen, Grebel, \& Harbeck 2003 for
counterarguments), this would also allow them to more easily survive 
reionization.  

Although low-mass galaxies continue to pose problems 
for CDM galaxy formation models, they remain important windows into 
understanding this process.  However, the 
opportunity to extend optical measurements of MSTO ages
throughout the Local Group (e.g., Brown et al.\ 2003) will be lost
with the end of the Hubble Space Telescope.

\acknowledgments

EKG was supported by the Swiss National Science Foundation through
grant 200021-101924/1.
JSG acknowledges funding from NSF grant AST-9803018 to the University of
Wisconsin and from its Graduate School. We thank the referee for valuable 
comments. 
This research has made use of NASA's Astrophysics Data System.

\clearpage

\begin{figure}
\epsscale{.80}
\plotone{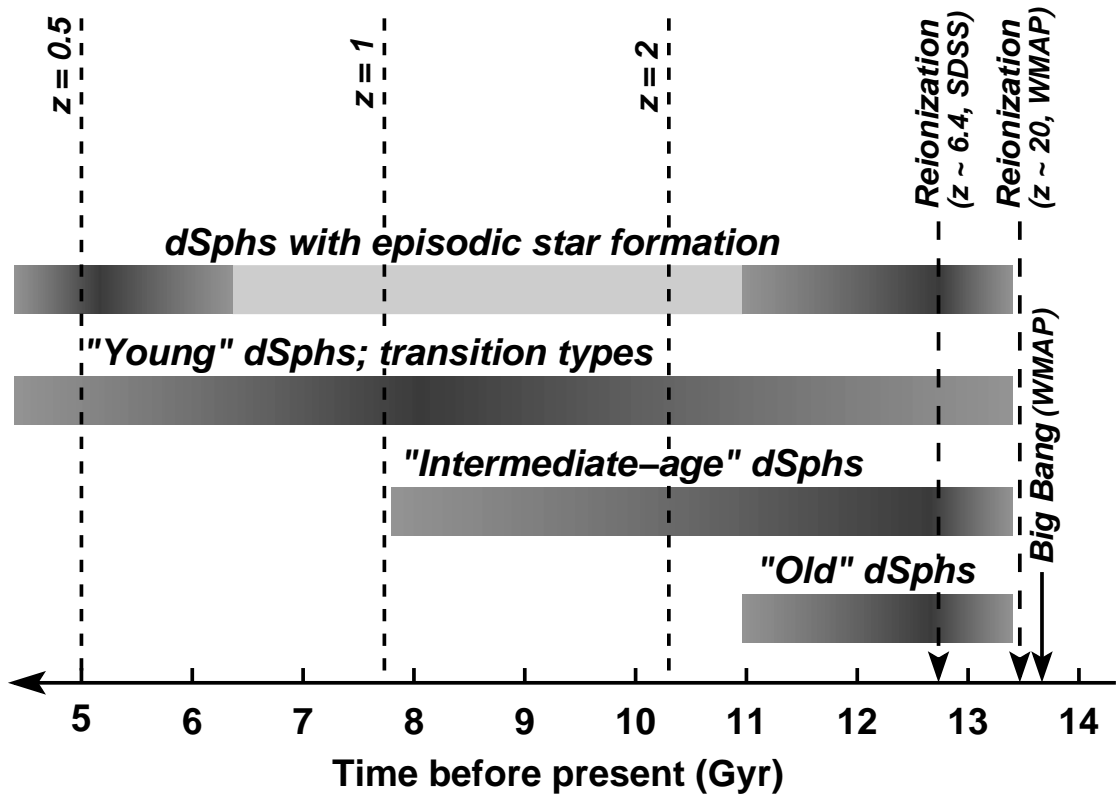}
\caption{
Sketch of the approximate duration of star formation 
episodes in dwarf spheroidal galaxies.  Darker shades indicate higher
star-formation activity.  Cosmological galaxy evolution
models predict a drop in or the cessation of star-formation activity
in low-mass dwarfs due to reionization, but this is not observed.
}
\end{figure}

\end{document}